# Spall strength of symmetric tilt grain boundaries in 6H silicon carbide

Chunyu Li and Alejandro Strachan

School of Materials Engineering, Purdue University, West Lafayette, IN 47906

## Abstract

Characterizing microstructural effects on the dynamical response of materials is challenging due to the extreme conditions and the short timescales involved. For example, little is known about how grain boundary characteristics affect spall strength. This study explores 6H-SiC bicrystals under shock waves via large-scale molecular dynamics simulations. We focused on symmetric tilt grain boundaries with a wide range of misorientations and found that spall strength and dynamical fracture surface energy are strongly affected by the grain boundary microstructure, especially the excess free volume. Grain boundary energy also plays a considerable role. As expected, low-angle grain boundaries tend to have higher spallation strengths. We also extracted cohesive models for the dynamical strength of bulk systems and grain boundaries that can be used in continuum simulations.

## 1. Introduction

Silicon carbide (SiC) is a high-performance ceramic material known for its exceptional hardness, thermal stability, and resistance to wear and corrosion [1,2]. These properties, combined with its high specific strength, make SiC a crucial material across various industries, particularly in defense applications for armor systems [3, 4, 5]. These armor systems are designed for ballistic protection and are normally subject to very high strain rate deformation. Thus, spall failure is among the most important phenomena. Understanding SiC spall failure is essential for improving its impact resistance and optimizing its use in protective armor systems [6, 7]. A better understanding of the failure mechanisms involving SiC spallation could help design optimally layered armor systems that minimize fragmentation and enhance safety.

Since the earliest observation of spall fracture by Hopkinson in 1913 [8], spall failure has been studied extensively and researchers have developed foundational theories of dynamic fracture, especially for metals and alloys [6, 7, 9]. Spall is a tensile failure process under shock conditions, involving both extremely short time and small length scales. Spall strength, defined as the maximum tensile stress during the spall process, is usually much higher than the ultimate strength measured under static loading. Spall failure is mediated by the nucleation of voids, when the local tensile stress in the spall zone reaches a critical value, followed by void growth and coalescence. Therefore, a correlation between dynamical and static strength generally exists, but the selection of materials for structures under shockwave loading cannot be simply based on static strength [[10]].

Dense SiC is needed for applications in protection against ballistic impact. Over the past decades, several manufacturing techniques were developed for dense SiC; they include hot-pressed sintering [[11]], pressureless sintering [[12]], spark plasma sintering [[13]], and infiltration reaction sintering [[14]]. All these techniques can achieve fine-grained microstructures, though at different costs. Experimental tests have indicated that spark plasma and hot-pressed sintering generally result in SiC with higher spall strength [[15, 16]]. It is well known that spall strength is sensitive to strain rate [[17]] and temperature [[18]], both environmental temperature and the temperature increase resulting from hypervelocity impact. Other than these two external factors, the main factors influencing SiC spall strength originate from its microstructure, such as grain size and porosity. Lower porosity usually enhances spall strength [[19, 20, 21]] but the effect of grain size is more nuanced. Depending on grain size, the grain boundary volume fraction in sintered SiC can be significant. Ultrafine-grained SiC with 100–500 nm grain size typically has 2–10 vol% grain boundaries. Coarse-grained SiC with few-micron grain sizes usually has about 1 vol% grain boundaries [[22]]. As is the case in metals, ceramics can follow the Hall-Petch relationship, which linearly relates the fracture strength or hardness to the inverse square root of the average grain size, over a range of grain size [[23]]. However, ultrafine or nanocrystalline ceramics, grain boundary weakening or amorphization can lead to grain boundary sliding and lead to the so-called inverse Hall-Petch relationship, i.e. softening rather than strengthening with grain size reduction [[24]]. Under shock conditions grain boundaries often facilitate the initiation and propagation of spall damage, especially when elevated temperature is inevitably involved in the shockwave propagation process. While experimental results show general trends of reduction in spall strength with decreasing grain size [[25], [26] , [27]], little is known about how specific grain boundary characteristics affect dynamical strength.

Molecular dynamics (MD) simulations have been extensively employed to understand material response to various loading conditions. As for the dynamic behavior of SiC, several studies focused on the failure of single crystals and nanocrystals [[28],[29],[30]]. Branicio et al. [[31]] reported on large-scale MD simulations of hypervelocity projectile impact on single crystal 3C–SiC using a many-body interatomic potential [[32]]. Because of their ultra-high particle velocity (15 km/s), atomistic damage mechanisms involving shock-induced structural transformation, plastic deformation, and brittle fracture and dynamic transitions between these regimes were observed. Li and co-workers [[18, 33, 34, 35]] studied 3C-SiC tensile fracture under shock conditions in recent years. They found that the spall strength of 3C-SiC has a considerable crystallographic dependence and a strong dependence on strain rate and temperature. The strain rate dependence can be fitted to a power law over a wide range of strain rates. For ultrafine-grain systems (diameter<32 nm) they found that the spall strength generally decreased with decreasing grain size, indicating an inverse Hall-Petch relationship. They also found grain boundaries have a strong negative effect on the performance of nanocrystalline 3C-SiC, lowering the spall strength to about 1/3 of single crystalline 3C-SiC. Feng and his co-workers [[36]] studied single crystalline 6H-SiC and found that intensive shock compressions cause amorphization and nanograin formation in hexagonal SiC due to limited availability of slip planes. Depending on the temperature and strain rate, the size of nanograins would be different and nanograin refinement occurs with increasing shock intensity.

Grain boundaries influence key material properties such as mechanical strength, fracture toughness and spall resistance. Polycrystalline and nanocrystalline SiC samples contain a large number of grain boundaries with varying characters and misfit angles. Here we characterize the spall strength of single GBs by studying bi-crystals with the GB situated at plane of maximum dynamical tension. There have been limited studies on the SiC bicrystals. Wojdyr et al. [37] developed a MD-based scheme to search for minimum energy structures of symmetric tilt GBs with different misorientations. Bringuier et al. [38] studied the dynamics of 3C-SiC bicrystals under shear by MD simulations using the well-known Tersoff bond-order potential [39]. It was shown that the GB sliding occurs at stress levels of about 20-30 GPa, with low-angle GBs exhibiting more pronounced sliding at low temperatures and high-angle GBs exhibiting dampened sliding due to atomic disordering. Gur et al. [23] conducted MD simulations on 3C-SiC bicrystals of various orientations also using the Tersoff bond-order potential and calculated material properties such as elastic modulus, fracture strength, and fracture toughness. Their simulation show a normal distribution of fracture strength with an average value ~26 ($\pm$9) GPa and a lognormal distribution of fracture energy with an average value 2.5 ($\pm$0.9) J/m$^2$. Guziewski et al. [40] extended a single-element Monte Carlo GB optimization approach to SiC system and were able to generate a large set of symmetric tilt or twist GBs in 3C-SiC as well as datasets of GB energy at low computational cost. Guziewski and coworkers [41, 42] further developed machine-learning models based on these GB datasets for identifying the importance of microscopic and macroscopic parameters in predicting GB properties in 3C-SiC including GB energy and tensile strength.

The objective of this study is to explore 6H-SiC bicrystals with various misorientations under shock conditions via large scale MD simulations. Simulations were conducted by using the open source code LAMMPS [43] with an interatomic potential developed by Vashishta et al. [32] for SiC. One of our focuses is to investigate the spall strength of various GBs with different misorientations and reveal the correlations between the spall strength and other GB properties. Another is to establish GB cohesive models and obtain dynamical fracture energy of various GBs under shock conditions and provide this information for the use in mesoscale simulations.

## 2. Methods

**Model system and symmetric tilt boundary generation**. There are two commonly used approaches to generate shock waves in atomistic simulations. A momentum mirror can be used as an ideal piston with infinite mass. In this setup, a target material is launched into the fixed piston as the desired particle velocity. A second approach is the flyer-target method: in this case, both flyer plate and target systems are explicitly modeled using MD. After thermalization, the flyer and target systems are assigned a relative velocity and a collision is simulated using adiabatic MD (NVE ensemble). The flyer and target are of equal cross-sectional area with periodic boundary conditions in the transverse directions. In our case, the flyer is taken as half the length of the target in order to induce maximum dynamic tension approximately half-way through the target. To keep the center-of-mass velocity of the flyer-target system zero, the flyer and target are respectively assigned with initial velocities of $4/3\,u_p$ and $-2/3\,u_p$, as shown in Figure 1. Here $u_p$ stands for particle velocity and we focus on $u_p$ =1.5km/s in this study. The shock wave propagates in z direction, and the impact plane is in $xy$ plane.

Grain boundaries are interfacial regions of disrupted atomic order between perfect crystalline regions with different orientations. The thickness of grain boundary varies between 1 nm and 5 nm, and the GB atomic structure is often partially amorphous. From the viewpoint of crystallography, five degrees of freedom are required to describe a GB: three orientation-related DOFs and two geometric DOFs for the boundary plane, i.e., rotation axis $\vec{o}$ (2 DOFs), rotation angle $\theta$ (1 DOF), and normal vector $\vec{n}$ to the GB plane. Depending on the geometrical alignment between $\vec{o}$ and $\vec{n}$, there are tilt GB ($\vec{o} \perp \vec{n}$) and twist GB ($\vec{o} \| \vec{n}$). The relative rotation angle of each grain defines symmetric GB (rotating each grain equally in opposite directions) and asymmetric GB (rotating each grain unequally in opposite directions). Here we only focus on symmetric tilt GB without an amorphous layer.

Several software tools are available for constructing GB, for examples, GBstudio [44], CrystalMaker [45], Aimsgb [46], ASE [47], Atomsk [48] etc. We used Atomsk and one example of GB construction is demonstrated in Figure 1. Detailed construction procedure is included in the supplemental material. Table S-1 lists the dimensions of bicrystal SiC target, while the flyer is just half length of the target.

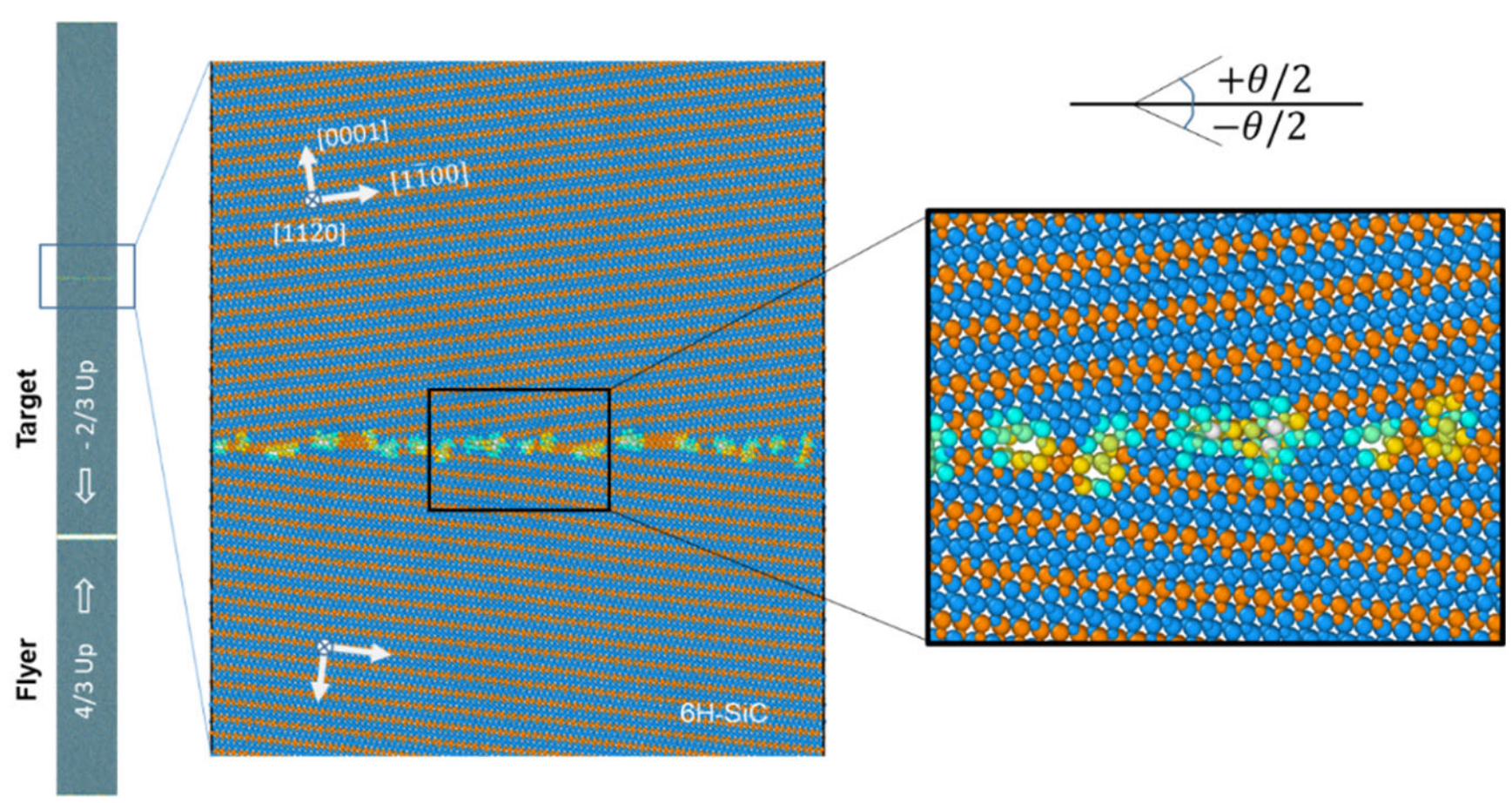

**Figure 1** Shock of bicrystal 6H-SiC with a symmetric tilt grain boundary (STGB)

**MD simulations of spall failure**. In our simulation setup, the target is represented by the bicrystal 6H-SiC generated above with a GB at the exact middle (as shown in Figure 1). In order to induce dynamic tension states in the GB vicinity, the target/flyer ratio has to be taken as 2. Therefore we cut off one half of the target as the flyer. The flyer and target are thus with exactly equal cross-sectional areas. The initial separation between the flyer and the target is 2 nm. While a free boundary is employed in the shock direction, periodic boundary conditions are used in the transverse directions. After 200ps canonical ensemble (NVT) relaxation, the flyer and target are assigned initial velocities of $4/3u_p$ and $-2/3u_p$, respectively. Here, $u_p$ refers to particle velocity. While this assignment is equivalent to the two plates impact against each other with +/- $u_p$, it results in zero velocity of the center-of-mass. We focus on particle velocity 1.5 km/s. Our previous simulations on 6H-SiC single crystals indicated spallation happens when particle velocity is above 1.2 km/s. The shock simulations are performed by using the microcanonical ensemble (NVE) for 40 ps with a time step 1.0 fs.

**Spall zone size.** The impact between the flyer and the target leads to the propagation of two compressive stress waves in opposite directions. The shock waves induce rapid compression of the crystalline structure of SiC. When the shock waves reach the free surfaces, they are reflected back as rarefaction waves. The interaction between the two rarefaction waves induces a tensile loading in the target and causes spall failure when a critical stress state is reached. To capture the stress wave propagation history, one-dimensional binning along the shock direction is necessary. The bin size can be chosen differently, but here we choose 1 nm. The calculation of the atomic stress tensor ($\sigma_{\alpha\beta}$) is done using the following stress definition [49]:

$$\sigma_{\alpha\beta} = -\frac{1}{\Omega_a}\left[mv_\alpha v_\beta + \frac{1}{2}\sum_{n=1}^{N_p}\left(r_{1\alpha}F_{1\beta} + r_{2\alpha}F_{2\beta}\right) + \frac{1}{3}\sum_{n=1}^{N_a}\left(r_{1\alpha}F_{1\beta} + r_{2\alpha}F_{2\beta} + r_{3\alpha}F_{3\beta}\right)\right] \quad (1)$$

where $\Omega_a, m,$ and $v$ are the atomic volume, atomic mass and the velocity of atom, respectively. The first term is the kinetic energy contribution after removing the binned center-of-mass translational velocity. The second term is the pairwise energy contribution with $n$ looping over the $N_p$ neighbors of the atom of interest, and the third term is the contribution from three-body interactions with $n$ looping over $N_a$ three-body interactions that involve the atom. $r$ and $F$ represent the atom position and the corresponding force on the central atom and those involved in the specific interaction. It should be pointed out that the atomic volume is not a fixed value but an instantaneous variable adjusted by the bin volume divided by the number of atoms inside a bin.

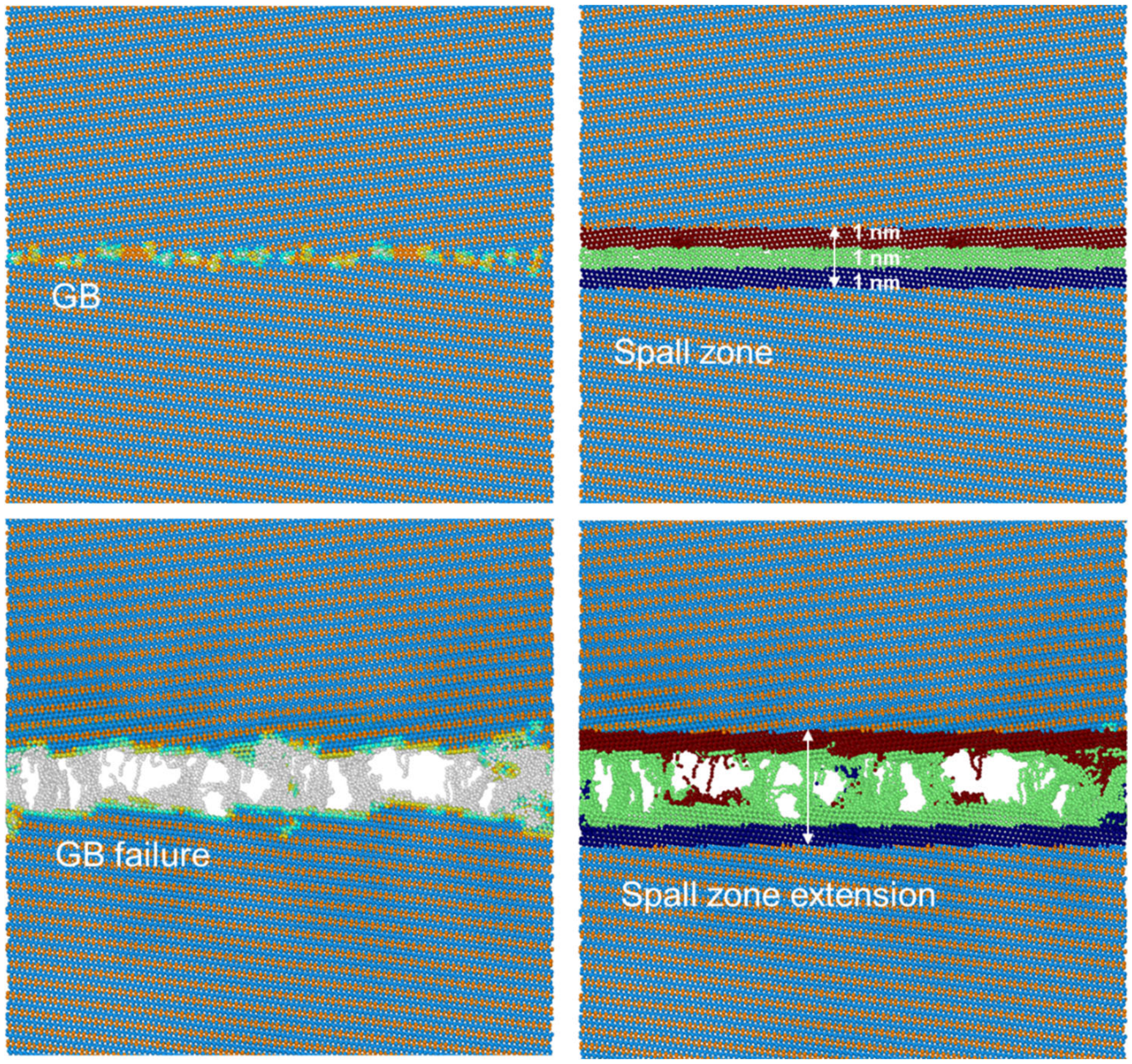

**Figure 2** Spall zone at the vicinity of GB

Spall failure usually begins with the nucleation of voids or from existing nano-porosity at the GB; void growth and coalescence eventually leads to a crack with rough surface along a plane parallel to the flyer impact surface. The size of spall zone depends on the particle velocity. Figure 2 shows the typical spall zone in our flyer-target simulations for $u_p$ = 1.5 km/s. One objective of this study is to obtain the cohesive model for the GB. Therefore, for a better calculation of local strain, we take a few extra bins surrounding the crack as the spall zone, mostly about 3 nm before deformation. We also calculate the local stress as the average atomic stress of the spall zone.

## 3. Shock wave propagation, spall, and dynamical traction law

As discussed above, we used a flyer/target length ratio of 1:2. The GB location is about 100nm from the impact plane. Figure 3(a) displays a typical history of local stress at the GB spall zone (~3nm). The local stress is the average atomic stress of the spall zone. The stress in the shock direction is clearly higher than those in another two directions, but all stresses in three directions follow the same trend. The stress wave takes a little over 10 ps to reach the GB and the maximum compressive stress in the shocking direction quickly increases up to about 50 GPa. The compression status stays for about 20ps. The compressive stresses then begin dropping and eventually tensile stresses appear. The tensile stresses increase rather quickly and reach peaks within about 0.5 ps. We take the maximum peak as spall strength.

The local strain calculation depends on the selection of base size of spall zone in the shock direction. There are two options: one is the initial size of spall zone at time 0; another is the size of spall zone at zero tensile stress when the stress changes from compression into tension. The difference between these two choices is actually very small, as shown in Figure 3(b). In order to get a better representation of stress-strain relationship at the moment of spallation, we choose to use the spall zone size at nearly zero tensile stress as the base length for calculating local strain. The strain rate corresponding to the moment of peak stress is calculated by 3-points linear fitting over the strain data around the time of peak stress, as demonstrated in the inset of Fig. 3(b), and the value is in the range from 8.5 x$10^{10}$ $s^{-1}$ to 2.5 x$10^{11}$ $s^{-1}$ for the various cases in this study.

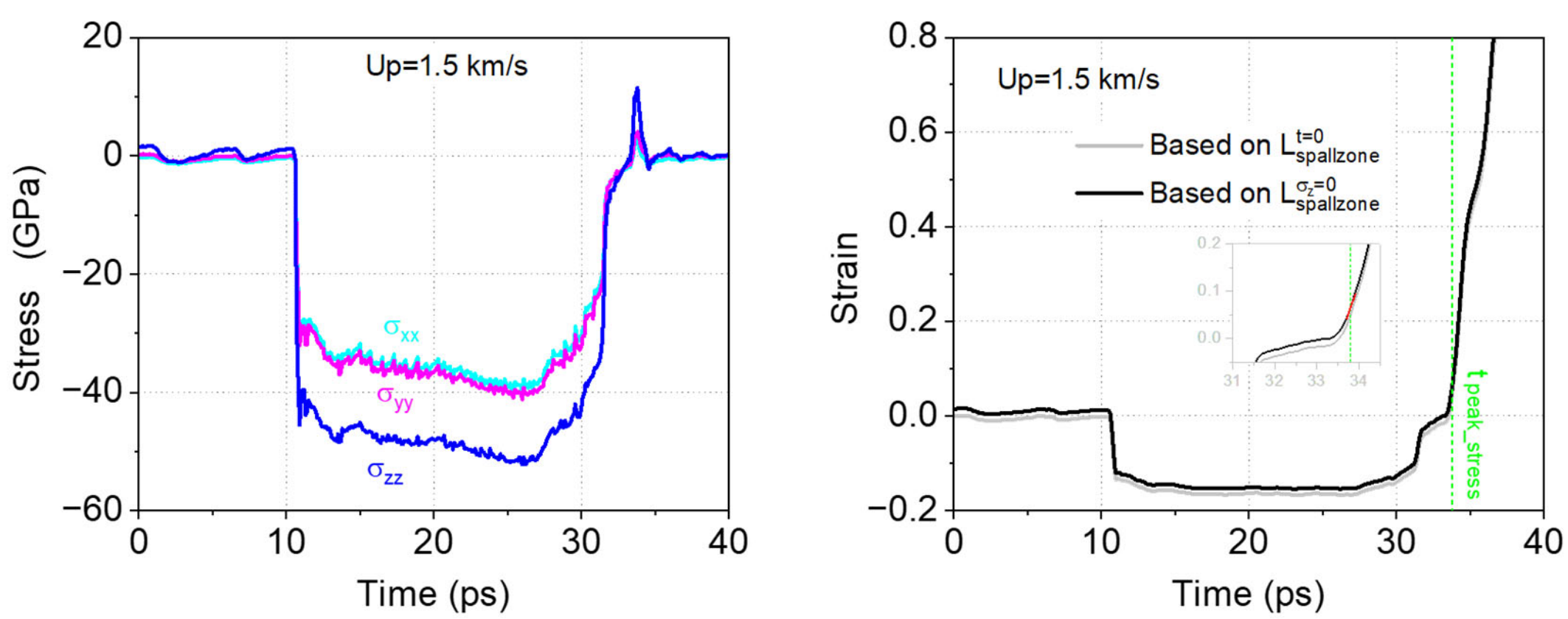

(a) (b)

**Figure 3** History of local stress and strain at the spall zone: (a) stress; (b) strain (inset showing the calculation of strain rate)

**GB cohesive model and dynamical fracture surface energy**. The stress-strain relationship for the GB spall zone can be readily extracted from the local stress and local strain histories mentioned above. From the stress shown in Figure 3(a), we can see that there are some fluctuations in the stress after spallation due to the nature of MD simulations. But we are only interested in the stress-strain relationship around the first peak when spallation is happening. Figure 4 shows the exact stress-strain relationship at the GB spall zone during the time frame of spallation. This dynamic stress-strain relationship is significantly different from that resulted from a longitudinal stretching (shown in Fig. 4(a) inset), in which the peak stress is about 7.15 GPa and failure strain is about 0.17, though the strain rates in both circumstances are in the same order of $10^{10}$ $s^{-1}$.

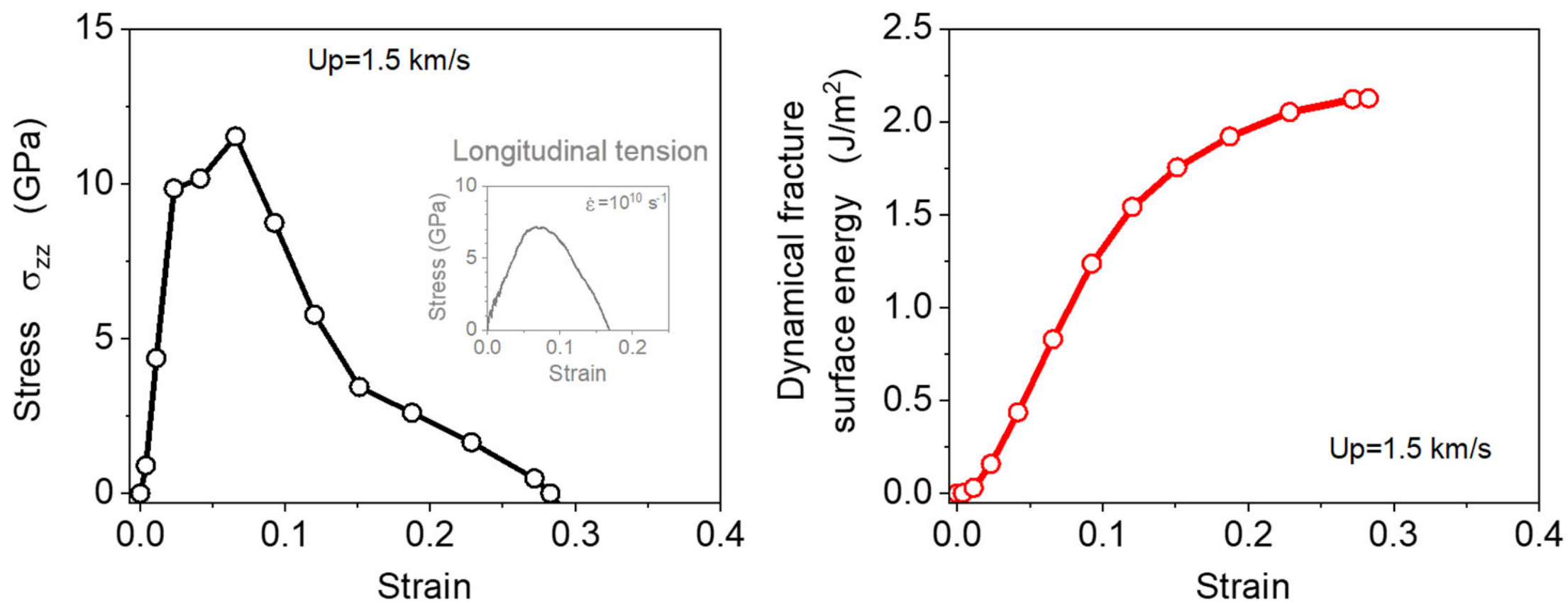

**Figure 4** GB spall zone: (a) Stress-strain relationship and (b) dynamical fracture surface energy

With a stress-strain relationship available, we can readily obtain the strain energy density $U$, i.e.,

$$U = \int \sigma d\varepsilon, \tag{1}$$

Where $\sigma$ is the longitudinal stress and $\varepsilon$ is strain. The work of crack opening can be expressed as:

$$W = \int F ds, \tag{2}$$

Where $F$ is the traction force at crack surface and $s$ is crack opening displacement. If we consider a rather perfect crack with a smooth surface, the dynamical fracture surface energy $\gamma$ can then be obtained as:

$$\gamma = W/2A, \tag{3}$$

Where A is the cross-sectional area of crack surface and a crack has both upper and lower surface. Considering stress $\sigma = F/A$ and displacement $s = \varepsilon L_0$ with $L_0$ the initial thickness of the spall zone, we define a dynamical fracture surface energy as:

$$\gamma = \frac{1}{2} L_0 \int \sigma d\varepsilon. \tag{4}$$

Based on the stress-strain curve shown in Fig. 4(a), and the size of GB spall zone ($L_0$=3 nm), we calculated the evolution of the dynamical fracture surface energy from Eq. (4) and the result is shown in Fig. 4(b). The dynamical fracture surface energy increases during the tensile process until a critical point (in this case the critical value is about 2.0 J/m$^2$), at which a spallation failure occurred.

## 4. Role of tilt grain boundary on spall strength

We have conducted MD simulations on a total of 11 STGB systems with various misorientation angles. Figure 5(a) shows the stress-strain relationship in the region around the spall plane. Initially, the stress increases approximately linearly with strain up to a maximum, the spall strength, within a small strain ranging from 0.05 to 0.08. The stress abruptly drops down during failure. In some cases the stress drops to zero, indicating no load transfer and in some cases, especially lower angle GBs, failure occurs more gradually over an extended strain. Figure 5(b) shows the corresponding stress-displacement curves, typical of cohesive models, stressing the relatively small displacements achieved during the dynamical failure of brittle materials.

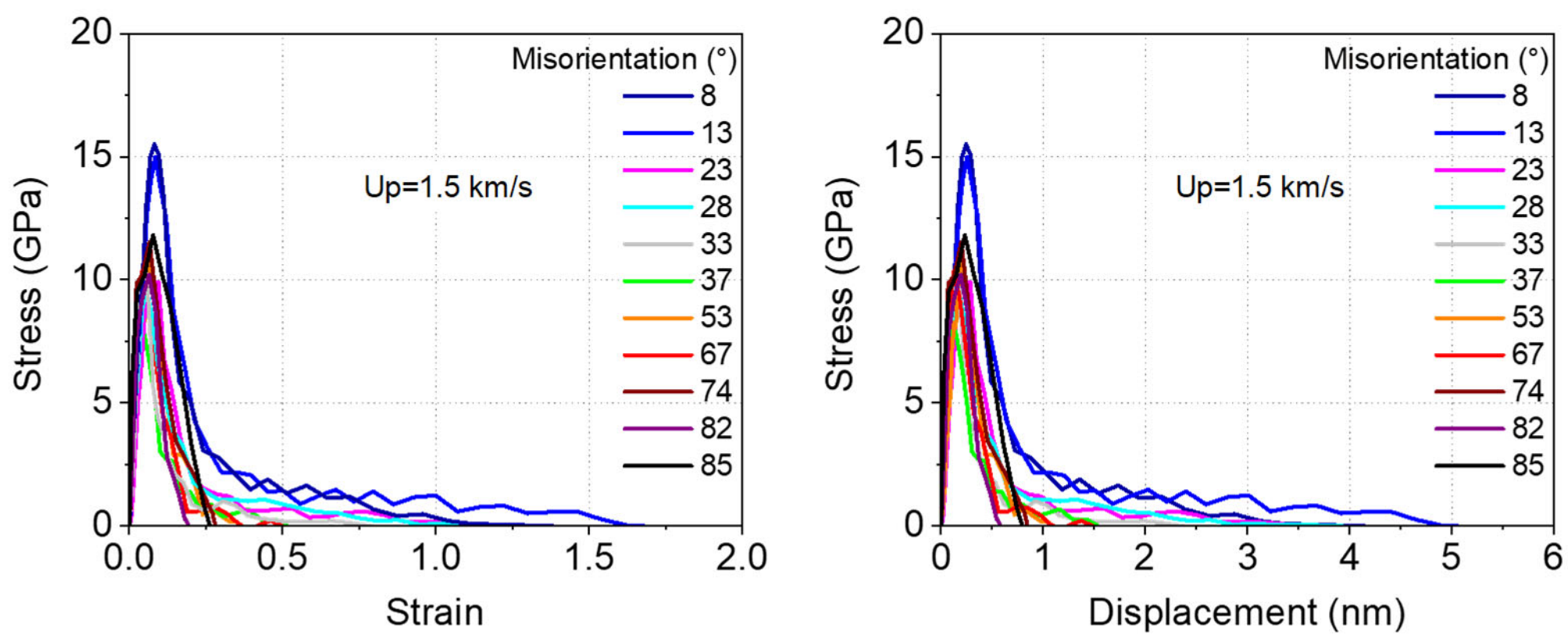

(a) (b)

**Figure 5** GB cohesive model: (a) Stress-strain relationship and (b) Stress-displacement relationship

Figure 6(a) shows the dependence of the GB spall strength, taken from the peak of Fig. 5, on the misorientation angle. Clearly, low-angle GBs exhibit higher spall strength and for high-angle GBs, the spall strength shows a rather constant value around 10GPa. To explore the effect of dynamical loading on GB strength, we subjected our samples to uniaxial strain using MD simulations for all 11 STGB systems with GB in the middle plane and periodic conditions in three directions. The strain rate used was $10^{10}$ $s^{-1}$. The correlation between the tensile strength and the spall strength is shown in Fig. 6(b). As expected, the tensile strength is lower than the spall strength and we find a correlation between the two measures of strength. The high-angle mismatch cases that result in lower strength show no correlation, indicating stochastic noise in the simulations.

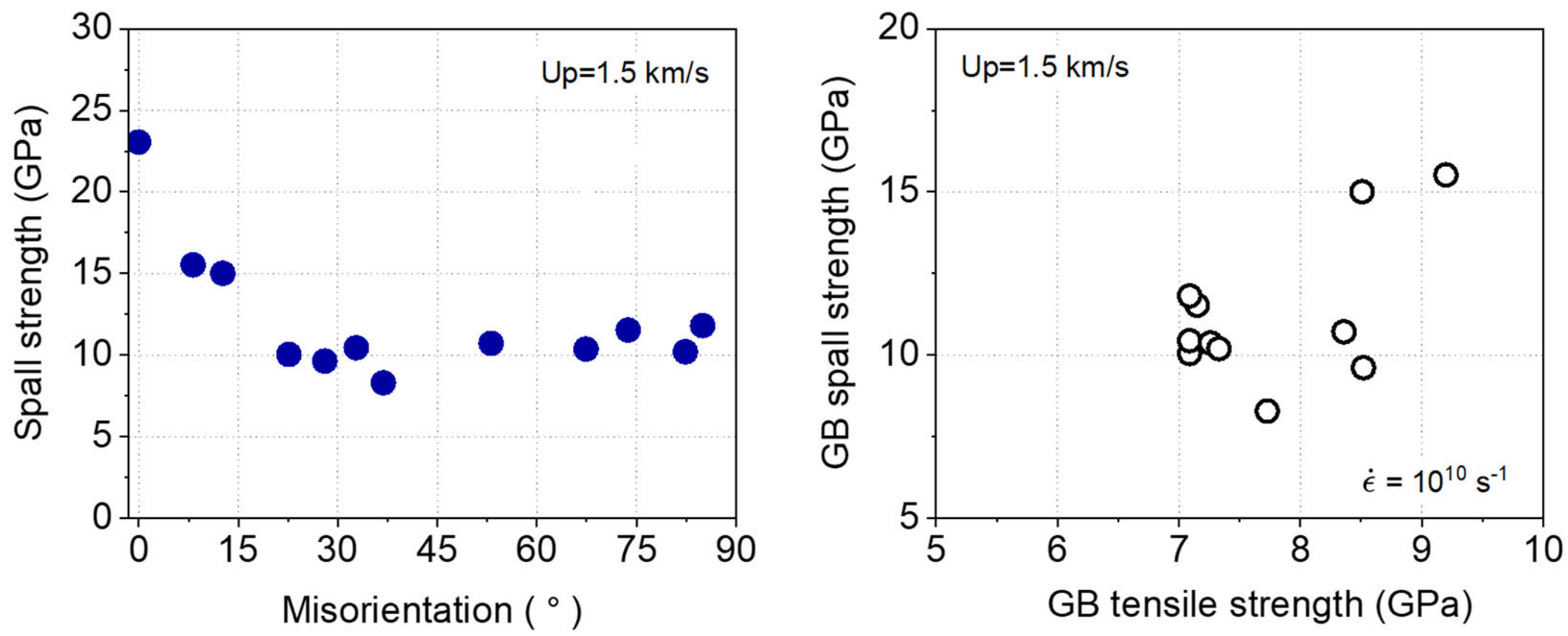

(a) (b)

**Figure 6** (a) Dependence of GB spall strength on misorientation and (b) correlation between spall strength and tensile strength

**Dynamical fracture surface energy.** Figure 7(a) shows the cumulative variation of dynamical fracture surface energy during the spall failure process of bicrystal GB with various misorientation angles. Figure 7(b) shows the dependence of critical fracture surface energy on misorientation angle. The critical fracture surface energy is in the range of 1.5 J/m$^2$ to 5.5 J/m$^2$. GBs with misorientation angles less than 30 degrees generally have a higher critical energy and their spall failures occur at a considerably larger breaking strain, which means a higher toughness. GBs with misorientation angle larger than 30 degrees have roughly the same critical energy and the breaking strain varies in a range but relative small.

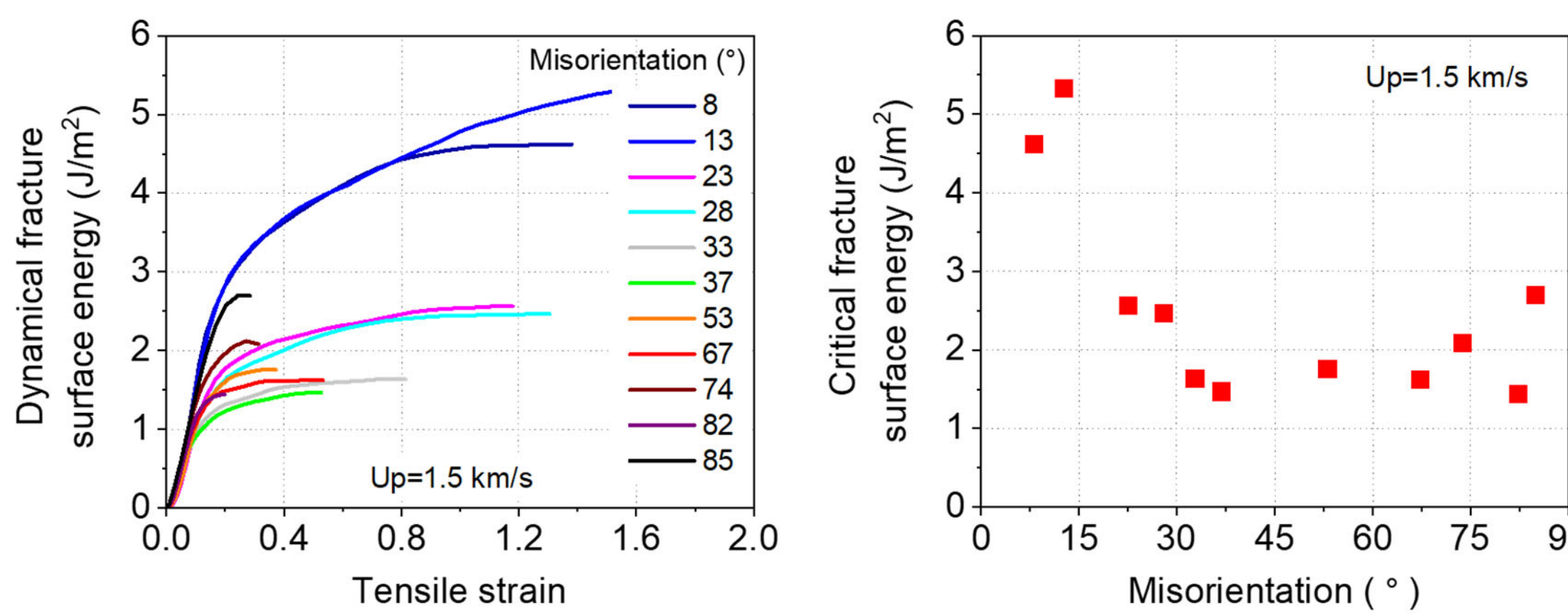

(a) (b)

**Figure 7** Dynamical fracture surface energy: (a) Cumulative fracture surface energy during spall failure; (b) Dependence of critical fracture energy on GB misorientation

**Effect of excess volume on GB spallation**. One fundamental GB microstructural parameter is the excess free volume (EFV). It is a measurement of local volume expansion due to GB formation. Excessive free volume tends to weaken GB, making it a preferred site for fracture initiation. It has been reported that EFV plays a dominant role in the tensile strength of SiC metastable GBs [42] so we studied its effect on spall strength. The EFV simply defined as:

$$V_{ex} = \frac{V_{total} - N\, V_{SiC}}{2A_{GB}}, \quad (5)$$

Where $V_{total}$ is the total volume of a bicrystal system, $V_{SiC}$ is the volume of a Si-C pair in a perfect 6H-SiC crystal, and $A_{GB}$ is the cross-section area of the GB. The factor of 2 accounts for the two GBs in our periodic bicrystal simulation setup. Here $V_{ex}$ is actually a normalized EFV with a length unit (Å$^3$/Å$^2$).

The EFV values for different GBs at 300K are shown in Fig. 8 (the error bars are from 200ps relaxation of the GB systems). As expected, GB with a smaller misorientation angle tends to have a smaller EFV, though some fluctuations exist. If we plot spall strength and critical fracture surface energy against EFV, as shown in Fig. 9, we can clearly see the effect of EFV. The general trend is that a GB with a very smaller EFV (<0.5 Å) tends to have a higher spall strength and higher critical fracture surface energy. But the effect saturates for EFV larger than 0.5 Å.

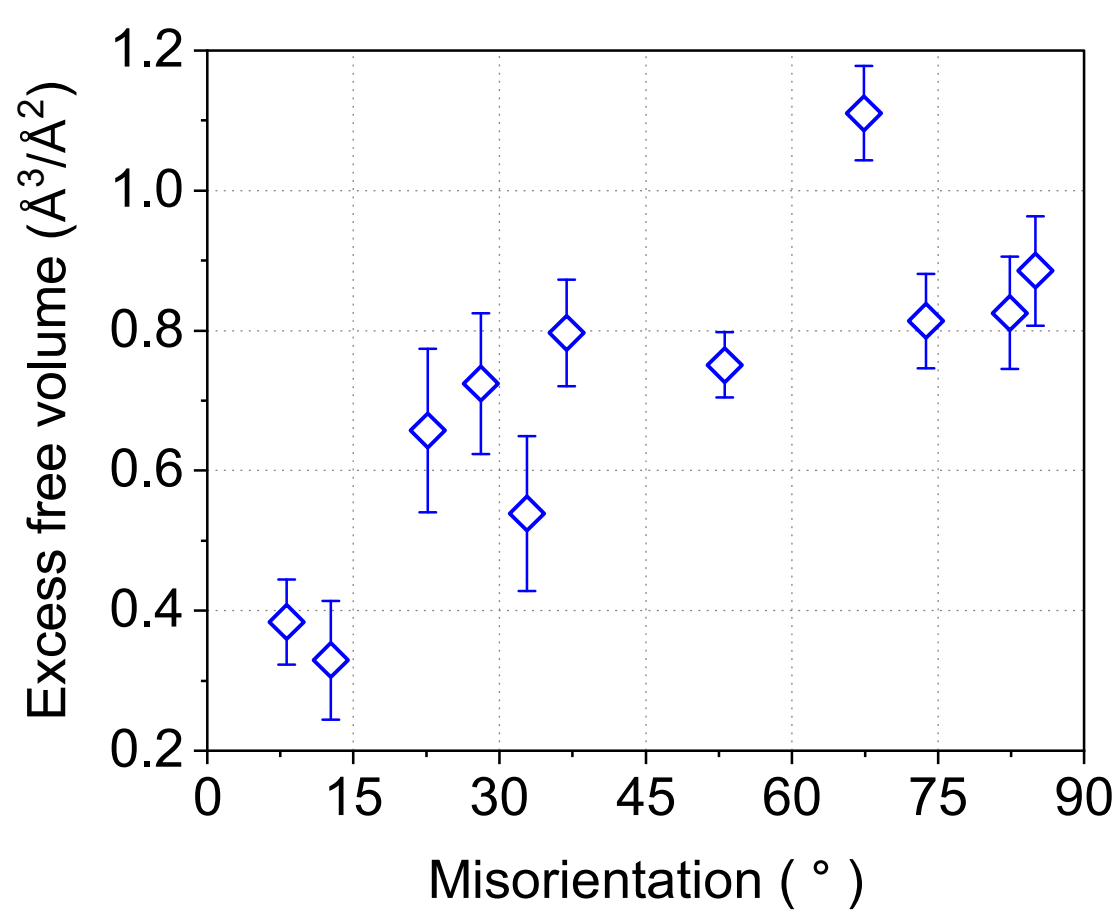

**Figure 8** Excess free volume as a function of GB misorientation

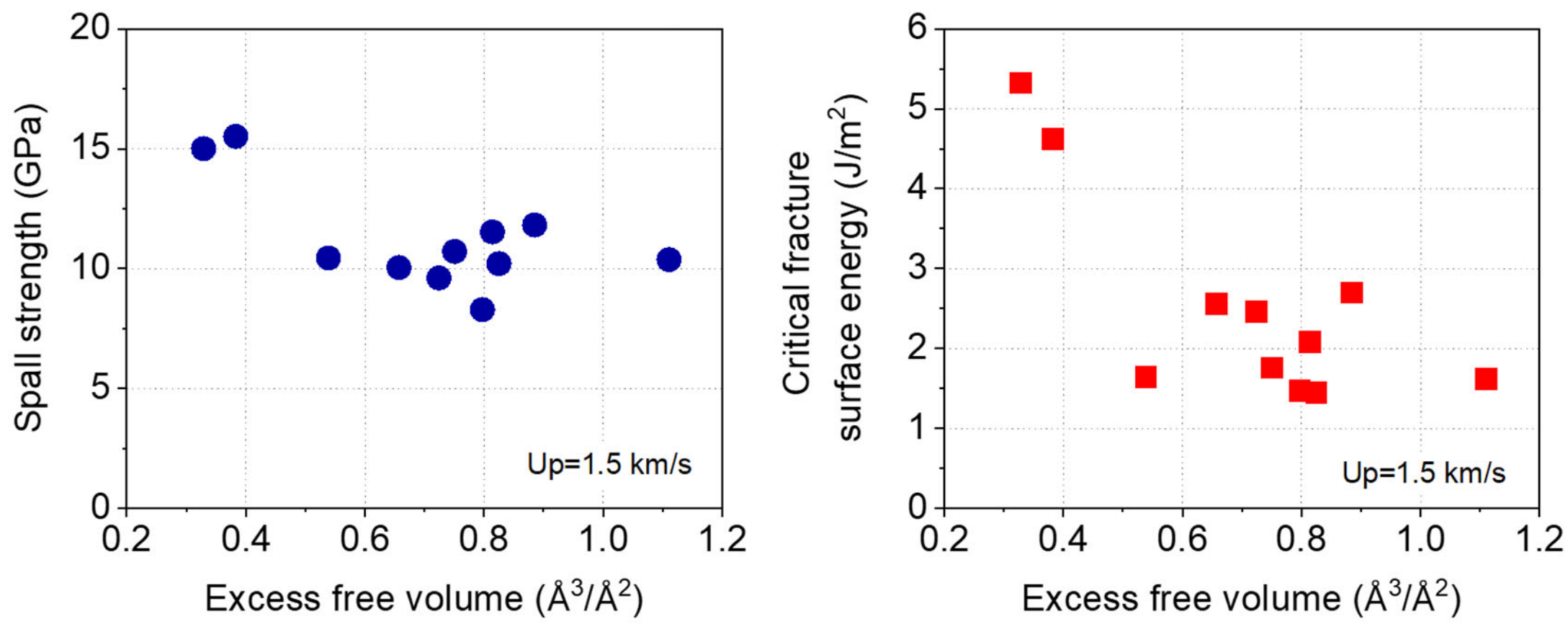

**Figure 9** Effect of excess free volume (a) on spall strength; (b) on critical dynamical fracture surface energy

**Effect of grain boundary energy on GB spallation.** Another important parameter related to GB structure is the grain boundary energy (GBE), which is the excess energy due to atomic mismatch at the interface between two grains. Generally, GBE is affected by misorentation angle, boundary microstructure, and impurity segregation. High-angle GB with a higher GBE is usually more susceptible to crack initiation, void formation, and intergranular fracture, leading to lower tensile strength [50, 51]. It has been reported that GBE is one of a few dominant indicators for predicting SiC tensile strength [42]. In this study, we are trying to reveal the correlation of GBE with the spall strength.

Similar to the EFV calculation, GBE $\gamma_{GB}$ can be calculated by the difference of potential energies between the GB system and perfect crystal system with the same number of Si-C pairs:

$$\gamma_{GB} = \frac{PE_{total} - N\,PE_{SiC}}{2A_{GB}}, \qquad (5)$$

Where $PE_{total}$ is the total potential enegy of a bicrystal system, $PE_{SiC}$ is the potential energy of a Si-C pair in a perfect 6H-SiC crystal. Usually, the potential energies are just taken as the minimum energy after a certain minimization process, which sometimes needs to be carefully designed [52, 53], especially for low-angle GBs. Here we are not trying to search the global minimum of a GB structure, instead we just take the average potential energy from the GB system and a perfect crystal system, both are well-relaxed at 300K under isothermal-isobaric NPT condition as mentioned before. The cross-sectional area is also an average value from the relaxation process.

Figure 10 shows the GBE as a function of misorientation angle for our 11 STGB systems. We find that the GBE depends on the misorientation angle and is in fair agreement with prior results, which are for 3C-SiC by MD simulations [52, 53]. We stress that our approach is not designed to find the minimum energy grain boundary which may or may not be realizable experimentally. To look into the correlations, we plot the spall strength and critical fracture surface energy against GBE, as shown in Figure 11. It seems to be that the correlation either between GBE and the spall strength or between GBE and the critical fracture surface energy is not significant for higher GBEs (>3.0 J/m$^2$). But for lower GBEs (<3.0 J/m$^2$), the spall strength or the critical fracture surface energy indeed considerably higher. Generally speaking, GBE is potentially a good indicator for the estimation of spall strength.

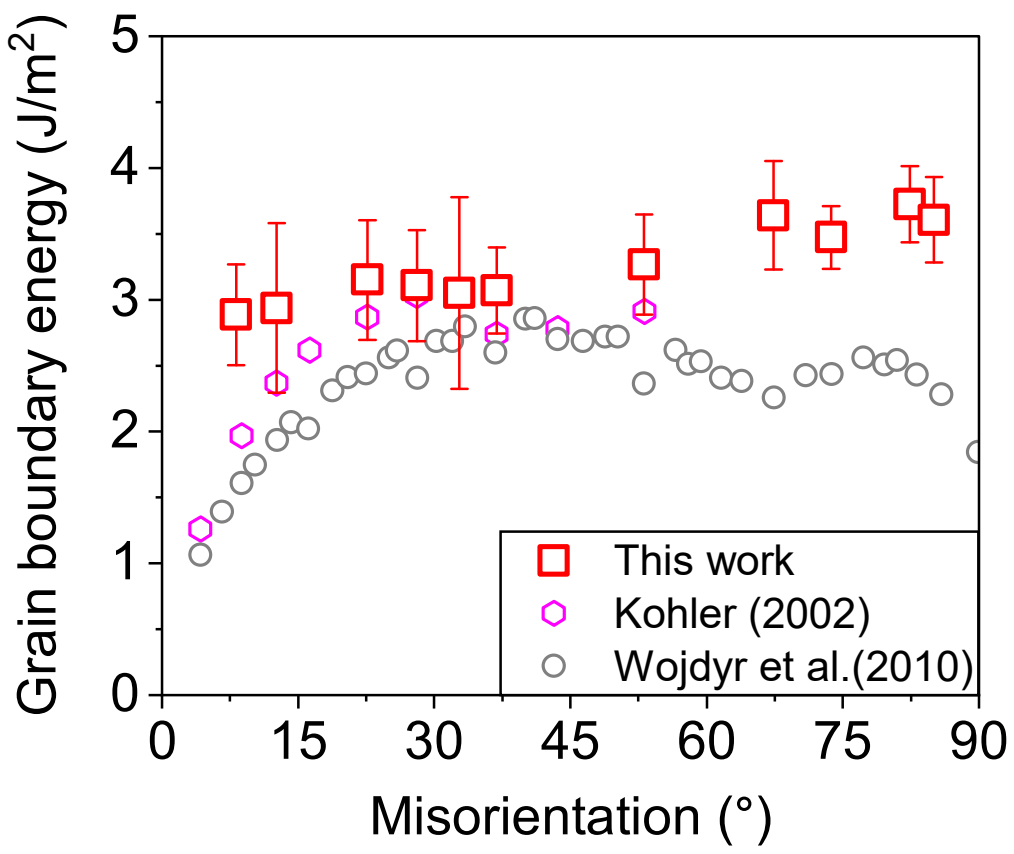

**Figure 10** Grain boundary energy (Note: Kohler (2002) and Wojdyr et al. (2010) results are for 3C-SiC [001] STGB)

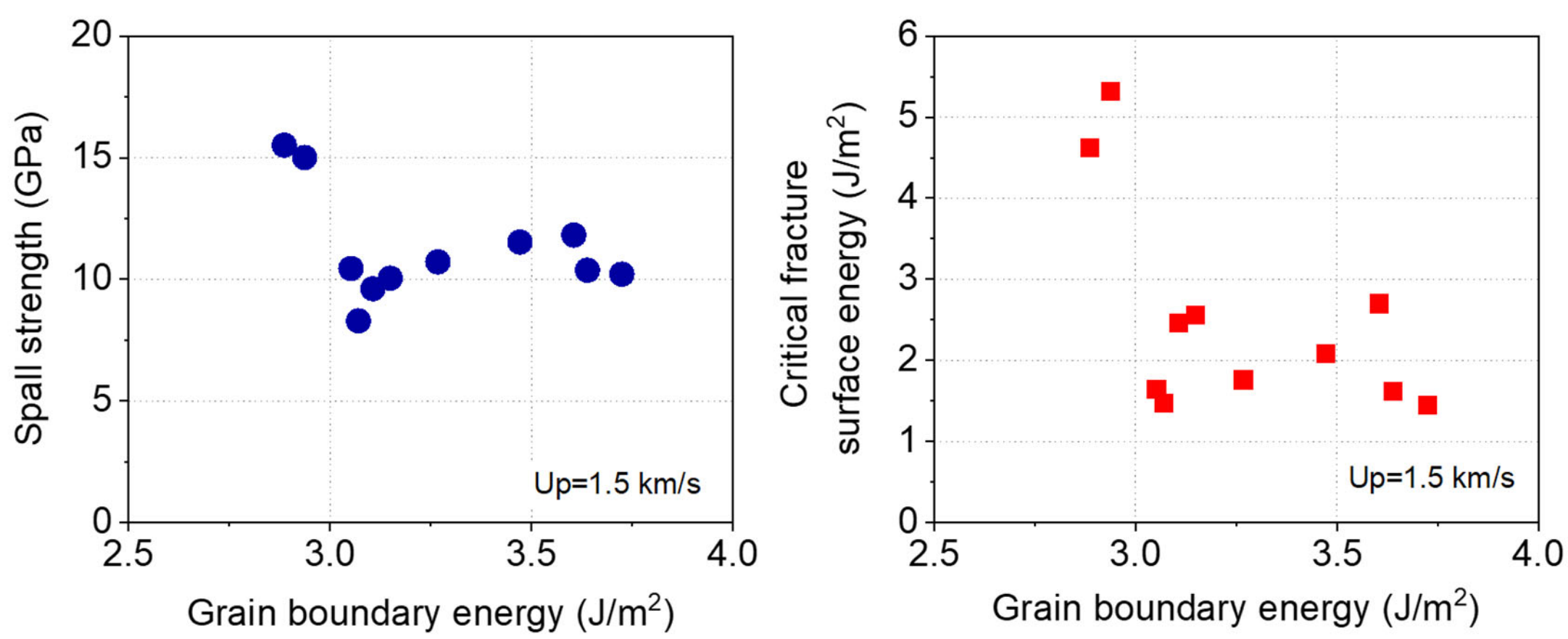

**Figure 11** Effect of GBE on spall strength and crtical fracture surface energy.

## 5. Discussion

From our simulation results, the spall strength is in the range 8-16 GPa under shock loading with particle velocity 1.5 km/s. This is consistent with prior MD simulations [33, 34, 35] but much higher than the experimentally determined spall strength, 0.5-1.5 GPa [15, 16]. Several reasons are behind this difference. Most importantly, spall strength is strongly strain-rate dependent. In shock experiments, the strain rate is usually much lower, in the order of $10^4$-$10^6$ s$^{-1}$, although higher strain rates are now achievable for laser shocks. The strain rate in our MD simulations is much higher, in the order of $10^{10}$ $s^{-1}$. This high strain rate is mainly due to the small dimension of our atomistic systems, which are in the order of 100 nm along the shock direction. Experimentally, the specimen size is no smaller that several hundreds of microns and often much thicker. Secondly, the GB microstructures in our bicrystal simulations are rather ideal, without impurities, significant amorphization, nor voids.

We also want to point out that the results shown in this study focus on GB properties. The simulations of these 6H-SiC bicrystals have indicate the possibility of more complex phenomena. In some cases, we observe intragranular failure together GB spall. As expected, the spall strength based on the intragranular failure would be roughly the same as that of single-crystalline SiC. These phenomena will be discussed in a subsequent publication.

## 6. Conclusions

We studied SiC bicrystals with various misorientations under shock loading conditions using large-scale atomistic simulations. Our study focused on the spall strength of symmetric tilt grain boundary. The bicrystal setup isolates the effects of a single grain boundary, allowing us to focus on its interaction with shock waves without the complexity of polycrystalline structures. Simulation results indicate that spall strength and dynamical fracture energy are strongly affected by the grain boundary microstructure, most dominantly its excess free volume. A grain boundary with a smaller excess free volume tends to have a higher spall strength and higher critical dynamical fracture energy. Excess free volume at grain boundaries depends on processing conditions and the nature of the GB, with low misorientation cases having, on average, lower excess free volumes. Thus, low-angle grain boundaries tend to exhibit higher spall strengths and also higher critical dynamical fracture energy. Grain boundary energy also plays a considerable role and the general trend is similar: a lower angle grain boundary tends to lead to higher spall strength.

The dynamic failure of SiC bicrystals is brittle, with a localized spall zone only a few nanometers in thickness. This is particularly true for high-angle grain boundaries. For low angle grain boundaries, damage spans over a slightly extended zone. The traction laws extracted from our MD simulations for various misorientation angles can be incorporated as cohesive models into the mesoscale simulations that can capture scales beyond the reach of MD.

## Acknowledgements

Research was sponsored by the Army Research Laboratory and was accomplished under Cooperative Agreement Number W911NF-22-2-0121/0101. The views and conclusions contained in this document are those of the authors and should not be interpreted as representing the official policies, either expressed or implied, of the Army Research Laboratory or the U.S. Government.

[40] M. Guziewski, A. D. Banadaki, S. Patala, S. P. Coleman, Application of Monte Carlo techniques to grain boundary structure optimization in silicon and silicon-carbide. Computational Materials Science 182: 109771 (2020).

[41] M. Guziewski, D. Montes de Oca Zapiain, R. Dingreville, S. P. Coleman, Microscopic and Macroscopic Characterization of Grain Boundary Energy and Strength in Silicon Carbide via Machine-Learning Techniques. ACS Appl. Mater. Interfaces 13: 3311−3324 (2021). https://doi.org/10.1021/acsami.0c15980

[42] D. Montes de Oca Zapiain, M. Guziewski, S.P. Coleman, R. Dingreville, Characterizing the Tensile Strength of Metastable Grain Boundaries in Silicon Carbide Using Machine Learning. J. Phys. Chem. C 124: 24809–24821 (2020). https://pubs.acs.org/doi/10.1021/acs.jpcc.0c07590

[43] A. P. Thompson, H. M. Aktulga, R. Berger, D. S. Bolintineanu, W. M. Brown, P. S. Crozier, P. J. in 't Veld, A. Kohlmeyer, S. G. Moore, T. D. Nguyen, R. Shan, M. J. Stevens, J. Tranchida, C. Trott, S. J. Plimpton, LAMMPS - a flexible simulation tool for particle-based materials modeling at the atomic, meso, and continuum scales. Comp Phys Comm, 271: 10817 (2022). https://doi.org/10.1016/j.cpc.2021.108171

[44] H. Ogawa, GBstudio: a builder software on periodic models of CSL boundaries for molecular simulation. Mater. Trans., 47 (2006), pp. 2706-2710. https://doi.org/10.2320/matertrans.47.2706.

[45] D.C. Palmer, Visualization and analysis of crystal structures using crystalmaker software. Z. Kristallogr. Cryst. Mater. 230 (2015), p. 559-572. https://doi.org/10.1515/zkri-2015-1869.

[46] J.L. Cheng, J. Luo, K.S. Yang, Aimsgb: An algorithm and open-source python library to generate periodic grain boundary structures. *Comput. Mater. Sci.* 155, 92-103 (2018).https://doi.org/10.1016/j.commatsci.2018.08.029

[47] A. H. Larsen, J. J. Mortensen, J. Blomqvist, I. E. Castelli, R. Christensen, M. Du lak, et al., The Atomic Simulation Environment–A Python Library for Working with Atoms. J. Phys.: Condens. Matter. 2017, 29, 273002. DOI 10.1088/1361-648X/aa680e.

[48] P. Hirel, Atomsk: A tool for manipulating and converting atomic data files. Comput. Phys. Comm. 197 (2015) 212-219. https://doi.org/10.1016/j.cpc.2015.07.012

[49] A. P. Thompson, S. J. Plimpton, W. Mattson. General formulation of pressure and stress tensor for arbitrary many-body interaction potentials under periodic boundary conditions. J. Chem. Phys. 131 (15): 154107 (2009). https://doi.org/10.1063/1.3245303.

[50] S. Lynch, A review of underlying reasons for intergranular cracking for a variety of failure modes and materials and examples of case histories. Engineering Failure Analysis 100: 329-350 (2019). https://doi.org/10.1016/j.engfailanal.2019.02.027.

[51] C. Sénac, Mechanisms and micromechanics of intergranular ductile fracture. Int J Solids Structures 301: 112951 (2024). https://doi.org/10.1016/j.ijsolstr.2024.112951.

[52] M. Wojdyr, S. Khalil, Y. Liu, I. Szlufarska, Energetics and structure of ⟨001⟩ tilt grain boundaries in SiC. Modelling Simul. Mater. Sci. Eng. 18 075009 (2010). DOI 10.1088/0965-0393/18/7/075009

[53] C. Kohler, Atomistic Modelling of Structures of Tilt Grain Boundaries and Antiphase Boundaries in b-Silicon Carbide. Phys. Stat. Sol. B 234(2): 522–540 (2002).